\documentclass[pra,aps,showpacs,nofootinbib,preprint]{revtex4}

\begin{document}

\preprint{DFPD-05/A/21}

\title{Comments on Long-Wavelength  Backreaction and Dark Energy}  

\author{Edward W. Kolb}\email{rocky@fnal.gov}
\affiliation{Particle Astrophysics Center, Fermi
       	National Accelerator Laboratory, 
Batavia, Illinois \ 60510-0500, USA \\
       	and Department of Astronomy and 
Astrophysics, Enrico Fermi Institute,
       	University of Chicago, Chicago, Illinois \ 60637-1433 USA}
\author{Sabino Matarrese}\email{sabino.matarrese@pd.infn.it}
\affiliation{Dipartimento di Fisica ``G.\ Galilei'' Universit\`{a} di Padova, 
        INFN Sezione di Padova, via Marzolo 8, Padova I-35131, Italy}
\author{Antonio Riotto}\email{antonio.riotto@pd.infn.it}
\affiliation{CERN, Theory Division, CH-1211 Geneva 23, Switzerland}

\date{\today}
\vskip 2cm
\begin{abstract}
In this brief note we comment on a recent attempt by Martineau and 
Brandenberger \cite{mb} to explain the acceleration of the Universe using the
back-reaction of {\em long-wavelength} perturbations associated with 
isocurvature perturbation modes. 
\end{abstract} 


\maketitle


In this brief note we would like to  comment on a work by Martineau and 
Brandenberger (MB)  \cite{mb}, who have recently presented a solution to the
dark energy problem making use of the gravitational backreaction of long
wavelength (super-Hubble) fluctuation modes on the background metric.  

The issue of the physical influence of the super-Hubble  modes on cosmological
observables,  such as the local Hubble expansion rate, has already been posed
in the literature \cite{mab,u,gb,fmvv,gb2,bl}. Sometime ago, some of us
\cite{mpla} have shown that if adiabaticity holds, then  {\it at any order} in
perturbation theory super-Hubble perturbations do not have any impact on local
physical observables (if gradients are neglected).  Let us briefly summarize
here the line of reasoning.  Our starting point is the Arnowitt-Deser-Misner
(ADM) formalism, with metric 
\begin{equation}
ds^2=-N^2\,dt^2+N_i\,dt\,dx^i+\gamma_{ij}\,dx^i\,dx^j . \nonumber
\end{equation} 

The three-metric $\gamma_{ij}$, and the lapse and the shift functions $N$ and
$N_i$, describe the evolution of the spacelike hypersurfaces.  In the ADM
formalism the equations simplify considerably if we set $N^i=0$. We may  then
express the spatial metric in the form
\begin{equation}
\gamma_{ij}=\exp\left[2\alpha(t,x^i)\right] \, h_{ij}(x^i) , 
\end{equation}  
where the conformal factor $\exp[\alpha(t,x^i)]$ may be interpreted as the
spatially dependent scale factor. The time-independent three-metric,
$h_{ij}(x^i)$, (with unit determinant) describes the three geometry of the
conformally transformed space. Within linear perturbation theory,
$\alpha(t,x^i)$ would contribute only to scalar perturbations, whereas $h_{ij}$
contains another scalar, as well as vector and tensor perturbations.  Vector
and tensor perturbations are necessary to satisfy Einstein equations beyond
linear order, however in the long wavelength limit they do not affect the
equations for the scalar perturbations.   Since $\exp[\alpha(t,x^i)]$ is
interpreted as a scale factor, we can define the local Hubble parameter
$H(t,x^i)=\dot{\alpha}(t,x^i)/N(t,x^i)$.

Now consider a Universe filled with  ideal fluid(s) described by an
energy-momentum tensor of the form $T_{\mu\nu}= \left(\rho+P\right)\,u_\mu
u_\nu +P\,g_{\mu\nu}$, where $\rho$ and $P$ are the energy  density and
pressure, respectively. The four-velocity vector can be chosen locally to be
$u^\mu=(1,\vec{0})$. In the case of a single fluid, this amounts to saying that
a volume measured by a local observer is comoving with the energy flow of the
fluid. In the multi-fluid case, the volume comoving with the total energy
density $\rho$ is not the same as the individual volumes comoving with the
individual energy density components. However, one can show that on
super-Hubble scales all comoving volumes become equivalent,  and therefore our
choice of the four velocity is well justified.

We consider the effect on the local expansion rate of the Universe and set
\begin{equation}
\label{psi}
\exp\left[\alpha(t,x^i)\right]=a(t)\,\exp\left[-\psi(t,x^i)\right] 
\end{equation}
and  choose to work in the synchronous gauge for which $N=1$ (together with
$N^i=0$). In this gauge the field equations look just like the familiar
homogeneous Friedmann-Robertson-Walker  equations point by point. As mentioned,
for a generic set of fluids we may safely take the four velocity to be
$u^\mu=(1,\vec{0})$ on super-Hubble scales, from which we define the local
expansion rate of tangential surfaces orthogonal to the fluid flow,
\textit{i.e.,} the local expansion rate, to be $\Theta\equiv  D_\mu u^\mu$. In
the synchronous gauge this quantity coincides with $3H$.

Using Eq.\ (\ref{psi}), we find that at any order in perturbation theory the
local expansion rate is $H(x^i,t)=\frac{1}{3}
\Theta(x^i,t)=\dot{a}/a-\dot\psi$.    We  infer that the local physical
expansion rate is influenced by long-wavelength perturbations only if the
gravitational potential is time dependent (if we  ignore spatial gradients in
$\psi$). Let us  analyze the case of an adiabatic fluid. From Einstein's
equations for a globally flat space we deduce (on super-Hubble scales) that
locally 
\begin{equation}
\label{j}
3H^2=8\pi G \rho =-8\pi G\,P + 2\dot H  .
\end{equation} 
If the pressure $P$ is a unique function of the energy density, we may expand
$\rho=\rho_0+\Delta\rho$, where $\rho_0$ is the energy density entering the
homogeneous Einstein equations from which the homogeneous scale factor $a(t)$
is computed. Inserting this expansion into Eq.\ (\ref{j}) and eliminating the
pressure in favor of the energy density $\rho\propto H^2$, we find a
differential equation of the generic form ${\cal
G}\left(\ddot\psi,\dot\psi\right)=0 $, which does not contain any term
proportional to $\psi$. Therefore, $\psi=\psi(x^i)$ is a solution of Einstein
equations {\it at any order} in perturbation theory and $\dot\psi=0$ holds on
super-Hubble scales.  From this generic argument we conclude that there is no
influence of very long wavelength modes on the physical local expansion rate of
the Universe if adiabaticity holds.  

In the adiabatic case, the influence of infrared modes is not locally
measurable. There is a simple explanation of this result. When adiabaticity
holds, the pressure is a well defined function of the energy density. This
means that the Hubble rate is only a function of the unique available physical
clock, the energy density. Indeed, suppose that the local expansion rate is
$H(t,x^i)=H(\rho(t,x^i),t)$.  This leads to
\begin{eqnarray}
\left(\frac{\partial H}{\partial t}\right)_{x^i}&=&
\left(\frac{\partial\rho}{\partial t}\right)_{x^i}
\left(\frac{\partial H}{\partial \rho}\right)_{t} +
\left(\frac{\partial t}{\partial t}\right)_{x^i}
\left(\frac{\partial H}{\partial t}\right)_{\rho}\nonumber\\
&=&-4\pi G(\rho+P) +\left(\frac{\partial H}{\partial t}\right)_{\rho}  ,
\label{mmm}
\end{eqnarray}
where we have made use of the equation $3H=-\dot\rho/(\rho+P)$. Comparing Eq.\
(\ref{mmm}) with $\dot H= -4\pi G(\rho+P)$, we see that $(\partial H/\partial
t)_{\rho}=0$, and hence $H(t,x^i)=H(\rho(t,x^i))$. The dependence of the local
expansion rate of the Universe on the clock time takes the same form as in the
unperturbed Universe when evaluated at a fixed value of the only clock
available, the energy density $\rho$.  Infrared modes do not have any locally
measurable effect on the expansion rate of the Universe. This result applies,
in particular, during inflation if the energy density is dominated by a single
inflaton field since in such a case adiabaticity is  guaranteed. One
implication of this result is that long-wavelength modes may not be responsible
for stopping inflation.

This conclusion does not hold when adiabaticity is not attained.  Indeed, the
right-hand side of Eq.\ (\ref{mmm}) would contain an extra factor
\begin{equation}
\left(\frac{\partial P}{\partial t}\right)_{x^i}
\left(\frac{\partial H}{\partial P}\right)_{t}  , 
\end{equation}
which makes it impossible for the local expansion rate to be a unique function
of the energy density.  The same conclusion was reached using a  second-order
approach \cite{gb,gb2,bl}.  

MB have very recently argued that one can make use of the fluctuations of a
scalar field $\varphi$ and so long as these fluctuations are associated with an
isocurvature mode, the effect on local observables should be measurable, so
that at late times the backreaction could mimic dark energy. This conclusion
is  reached by inspecting the second-order effective energy momentum tensor.
Their approach is not fully consistent as  the metric fluctuations and the
scalar field fluctuations are not properly expanded  at second-order to 
include intrinsically second-order quantities. However, for the sake of
comparison,  we will perform the same approximation.

Following Ref.\ \cite{mb}, let us   consider a scalar field $\varphi$ with
potential $V(\varphi)=\lambda\varphi^4$ . As it was shown in Ref.\ \cite{tra},
the equation of state of the zero mode $\varphi_0$ is that of radiation,
implying $\varphi_0\propto a^{-1}$, $a$ being the background scale factor.  The
scalar field  may be split as $\varphi=\varphi_0+\delta\varphi$ and one may
consider  the long-wavelength limit of the spatially averaged  energy density
of the scalar field. MB  argue that it  is dominated at large times by a piece
$\delta\rho_\varphi=\frac{1}{2}V^{\prime\prime}\langle\delta\varphi^2
\rangle=-\delta P_\varphi$. It dominates over the matter contribution, thus
leading to an effective cosmological constant behavior. The way this 
conclusion is reached is by making use of Einstein equations which relate the
scalar fluctuations to the metric fluctuations
\begin{equation}
\label{a}
(\dot{H}+3 H^2)\Phi=-4\pi G \left(V^\prime\delta\varphi+
\frac{\delta\rho_m}{2}\right) ,
\end{equation} 
where $\Phi$ is the metric (scalar) perturbation in the longitudinal gauge, and
dropping the piece proportional to $\delta\rho_m$. Since during a
matter-dominated epoch $H^2\sim a^{-3}\sim V^\prime$ and assuming that the
gravitational potential  $\Phi$ is constant,  MB conclude that $\delta\varphi$
does not evolve  with time, implying that  $\delta\rho_\varphi\sim -\delta
P_\varphi\sim \langle\Phi^2\rangle/a^2$. This term will eventually dominate
over  matter thus leading to acceleration. 

This result is, however suspicious because MB employ  the ``adiabatic'' part of
the fluctuation of the   scalar field, that is the part of the fluctuation
proportional to the gravitational potential. Let us work in the longitudinal
gauge with metric $ds^2=-(1+2\Phi)dt^2+a^2(t)(1-2\Phi)\delta_{ij}dx^i dx^j$.   
As we stressed above,  one needs a purely isocurvature component to hope that
super-Hubble modes have any influence on cosmological observables.  The
first-order Klein-Gordon equation for the fluctuation of the scalar field reads
\begin{equation}
\label{g}
\delta\ddot{\varphi}+3H\delta\dot{\varphi}+V^{\prime\prime}\delta\varphi=
4\dot{\varphi}_0\dot{\Phi}-2V^\prime\varphi_0\Phi\, .
\end{equation}
The exact  solution of this equation is (we assume a constant $\Phi$  as in
Ref.\ \cite{mb})
\begin{equation}
\label{exact}
\delta\varphi=q\frac{\delta\varphi_*}{\varphi_{0*}}\varphi_0+
t\dot{\varphi}_0\Phi\, ,
\end{equation}
where the label ``$*$'' indicates that the initial conditions have been fixed
at some  time $t_*$. The first piece represents the pure isocurvature component
of the fluctuation, while the piece proportional to the gravitational potential
is the adiabatic component. Its origin  can be understood via the separate
Universe approach: when the fluctuation starts having some nontrivial evolution
after inflation, initial conditions are different in different Hubble patches
if adiabatic perturbations set by the gravitational potential $\Phi$ are
present.  The parameter $q\sim 1$ depends upon the polynomial form of the
potential. From Eq.\ (\ref{exact}) we see that $\delta\varphi\sim \varphi_0\sim
a^{-1}$ throughout the all  evolution of the Universe. This means that
$\delta\rho_\varphi\sim a^{-4}$ during  all epochs and there is no way the
long-wavelength fluctuations of the scalar field can dominate over
nonrelativistic matter. 

The reason why MB reach a different conclusion is that they solve Eq.\
(\ref{a}) to infer the time behavior of the scalar fluctuation. However, the
piece proportional to $\delta\rho_m$ is dropped. This is somewhat inconsistent
since there are  extra components in that equation to account for the other
fluids (and indeed more than one fluid is necessary to have isocurvature
perturbation without which no super-Hubble  influence on observables is
possible). On the contrary, solving directly the Klein-Gordon equation shows
that super-Hubble fluctuations behave like a relativistic gas  as the zero
mode.

To understand this result in another way, we consider a nonlinear
generalization of the metric in the longitudinal gauge,
$ds^2=-e^{2\Phi}dt^2+e^{-2\Psi}a^2(t)\delta_{ij}dx^i d x^j$. Again, following
Ref.\ \cite{mb}, we take the two gravitational potentials equal, $\Phi=\Psi$
(even though this not fully  consistent), and constant in time (there is a mild
logarithmic dependence upon time at second-order, which is, however, not
relevant  in the considerations of Ref. \cite{mb}). As long as we are
interested in long-wavelength perturbations, we can  recast this perturbed
metric in a FRW form by putting $d\overline{t}=e^{\Phi}dt$ and
$\overline{a}=e^{-\Phi}a$.  The Klein-Gordon equation for the  quantum scalar
field containing the zero mode and the super-Hubble perturbations  reads 
\begin{equation}
\label{aa}
{\varphi}^{\prime\prime}+3\overline{H}{\varphi}^\prime+\frac{dV}{d\varphi}
=0 ,
\end{equation}

where the primes indicate now differentiation with respect to $\overline{t}$ 
and
$\overline{H}=(d\overline{a}/d\overline{t})= e^{-2\Phi}(H-\dot{\Phi})$.  Eq.\
(\ref{aa}) can then be rewritten as
\begin{equation}
\ddot{\varphi}+3H\dot{\varphi}+e^{2\Phi}\frac{dV}{d\varphi}=
4\dot{\varphi}\dot{\Phi}\, ,
\end{equation}
which expanded to first-order gives exactly the perturbed
Klein-Gordon equation, Eq.\ (\ref{g}). 

For a constant gravitational potential, one  concludes that the scalar field
(including the long-wavelength modes) behaves exactly like the zero mode (upon
rescaling: $\lambda\rightarrow e^{2\Phi} \lambda$). Therefore, the energy
density associated with the long-wavelength ``adiabatic''  scalar perturbations
scales like a relativistic gas (as the zero mode) and may never  dominate over 
matter.  Since super-Hubble perturbations may  have an effect on observables
only if they are associated with isocurvature perturbations, we conclude that
the effect, if any, must depend upon the isocurvature component of
$\delta\varphi$, {\it i.e.,} on  $\delta\varphi_{*}/\varphi_{0*}$. 
Indeed, it is
easy to see that the isocurvature perturbation
$S_{m\varphi}=3\left(\zeta_m-\zeta_\varphi \right)$, where the $\zeta_m$ and
$\zeta_\varphi$  are the comoving curvature perturbations for matter and the
scalar field respectively, is proportional to $\delta\varphi_*/\varphi_{0*}$.
However, isocurvature perturbations induce a sizable effect only when the
fluids have comparable energy densities, $\dot\zeta\sim
H\left(\dot{\rho}_\varphi\dot{\rho}_m/{\dot\rho}^2\right) S_{m\varphi}$, and it
is not clear to us if it suffices to have a super-Hubble  isocurvature
component to mimic dark energy.

\acknowledgments 

We thank D.\ H.\ Chung for useful correspondence.  A.R.\
is on leave of absence from INFN, Padova (Italy). E.W.K.\ is supported in part
by NASA grant NAG5-10842 and by the Department of Energy. S.M.\ acknowledges
partial financial support by INAF.



\end{document}